\documentclass[12pt]{article}
\usepackage{tikz}
\usetikzlibrary{arrows.meta,positioning}
\usepackage{amsfonts}
\usepackage{amsmath, amsthm, amssymb}
\usepackage{indentfirst}
\usepackage[left=1.2in,right=1.2in,top=1.2in,bottom=1.2in]{geometry}
\usepackage[natbibapa]{apacite}
\usepackage{multirow}
\usepackage{verbatim}
\usepackage{bbm}
\usepackage{scrextend}
\usepackage{lipsum}

\usepackage[colorlinks=true, linkcolor=red, citecolor=blue, urlcolor=blue]{hyperref}

\usepackage{bbm}
\usepackage{graphicx}
\usepackage{ragged2e}
\usepackage{booktabs}

\usepackage{mathtools}

\usepackage{float}
\usepackage{mathrsfs}
\usepackage{pdfsync}

\usepackage{pdfsync}

\usetikzlibrary{shapes,arrows}
\usepackage{graphics,color}
\usepackage{graphicx}
\usepackage{inputenc}

\newtheorem{theorem}{Theorem}

\DeclareMathOperator*{\argmin}{arg\,min}

\begin{document}

\title{Robust Market Design with Opaque Announcements}

\author{Aram Grigoryan\thanks{Department of Economics, University of California - San Diego, a2grigoryan@ucsd.edu} \and Markus M\"{o}ller\thanks{Department of Economics, University of Bonn, mmoelle2@uni-bonn.de}} 

\date{August 2024}

\maketitle

\begin{abstract}
We introduce a framework where the announcements of a clearinghouse about the allocation process are \textit{opaque} in the sense that there can be more than one outcome compatible with a realization of type reports. We ask whether desirable properties can be ensured under opacity in a robust sense. A property can be guaranteed under an opaque announcement if every mechanism compatible with it satisfies the property. We find an impossibility result: strategy-proofness cannot be guaranteed under \textit{any} level of opacity. In contrast, in some environments, weak Maskin monotonicity and non-bossiness can be guaranteed under opacity.
\end{abstract}

\section{Introduction}

Public announcements of clearinghouses often comprise a description of the allocation process (e.g., a pseudo-algorithm, a list of rules, or illustrative examples). Typically, such announcements aim to outline the outcomes one may encounter in different scenarios. Yet some details of the process often remain unspecified. For instance, Google did not entirely explain the determination of price floors in its online ads auctions and then deployed hidden programs that adjusted these floors in their favor.\footnote{Allegedly, Google used advertisers' current or historic bid observations without the publishers' or advertisers' knowledge to adapt the price floors and effective bids.}$^,$\footnote{\url{https://www.wired.com/story/google-antitrust-ad-market-lawsuit/}. Last accessed, July 2024.}

While opacity in announcements may be intentional or self-serving for the clearinghouse, another reason might just be that the process is complicated, poorly understood, or certain variables are unclear. Such scenarios may, for instance, occur in the context of reserves-based policies, which usually do not communicate the seat-processing order of reserves. In fact, Boston Public Schools faced controversies when their processing orders of reserved- and non-reserved slots were not specified. At the same time, the school district officials may not have been aware that this processing order affected the outcomes \citep{dur/kominers/pathak/sonmez:18}.\footnote{The lack of transparency and misconceptions in the reserves-processing order has been highlighted in other applications including US H-1B visa allocation \citep{pathak2022immigration} and admission to technical universities in India  \citep{sonmez/yenmez:19,aygun/turhan:22}.} Another common source of opacity is unspecified capacity in the context of matching applications such as college admission, school choice, and course allocation. Here, the exact numbers of school-,  college- or course seats are usually not communicated to applicants. Even the school districts or the colleges frequently do not know their final capacities \textit{a priori}. Interestingly, the capacities may also be adjusted within the admissions cycles depending on applicants' reports \citep{bobbio/carvalho/lodi/rios/torrico:23,dur2021capacity}.\footnote{For instance, the \textit{Course Match} program that assigns MBA students to courses \citep{budish/cacchon/kessler/othman:17} has used students' preference reports to cancel entire courses or increase course capacities \textit{on-line}, and re-run its algorithm to reoptimize the allocation for all students \citep{dur2021capacity}.}

In light of these observations, it seems interesting to ask whether specific desirable properties can still be assured under opacity. Indeed, some holistic statements about an institution (or process) may no longer be reliable in the presence of opacity.
For instance, as recently put by a Google employee about \textit{Dynamic Revenue Share (DRS)} and \textit{Reserve Price Optimization (RPO)} programs:\footnote{\url{https://www.wired.com/story/google-antitrust-ad-market-lawsuit/}. Last accessed, July 2024.}

\begin{addmargin}[1em]{2em}
\textit{``One known issue with the current DRS is that
it makes the auction untruthful. \dots{} Doesn’t that [RPO] undermine the whole idea of second-price auctions? It’ll transform the system into a 1st price auction where the bidder has a strong incentive to bid LESS than he’s willing to pay."}
\end{addmargin}

A central desirable property in various allocation problems is strategy-proofness. Several school districts using the \textit{Student-Proposing Deferred Acceptance} \citep{gale/shapley:62} essentially mention on their websites that their admission system is `strategy-proof'. The \textit{New York City Public School 2024 Admission Guides} \citep{nycpublicschools:24} writes:

 \begin{addmargin}[1em]{2em}
\textit{````... Place the programs on your application in your true order of preference, with your favorite program as \#1. There is no better strategy!''}
\end{addmargin}

A similar practice is common in other matching applications such as the Israel \textit{``Mechinot'' match} \citep{gonczarowski/kovalin/nisan/romm:19}, and \textit{National Resident Matching Program (NRMP)} \citep{nrmp:23}. Motivated by these examples, we take a robust approach to examine whether certain desirable properties, such as strategy-proofness, can be guaranteed under opaque announcements. 

We employ a general social choice problem where a central planner has to select an outcome for a set of individuals for each realization of reports (which we call a problem). The central planner makes an \textit{announcement}, which is a description of how problems translate into outcomes. The announcement may be \textit{opaque} in that the description may specify more than one possible outcome for some problems. We entertain the possibility that the central planner may induce any direct mechanism compatible with the description. Concretely, a mechanism is \textit{possible} if, for every problem, the mechanism's chosen outcome is in the set of possible outcomes specified by the announcement for that problem. Under this robust approach, an opaque announcement \textit{guarantees} a property if every possible mechanism satisfies the property.\footnote{We treat holistic statements about desirable properties of the central authority as pure cheap talk. Hence, in our framework, it is unnecessary to encompass holistic statements about desirable properties themselves. In fact, a clear disentanglement from the announcement is what allows us to better isolate the effect of opacity on the reliability of such holistic statements.}

To illustrate our opacity framework, consider a simple problem of allocating two objects, $1$ and $2$, to two individuals, Ann ($A$) and Bob ($B$). There are only two possible outcomes: let $x$ denote the outcome where $A$ receives object $1$ and $B$ receives object $2$, and let $y$ denote the outcome where $A$ receives object $1$ and $B$ receives object $2$. There are four potential problems, (1) both $A$ and $B$ prefer $x$ to $y$, (2) $A$ prefers $x$ to $y$, and $B$ prefers $y$ to $x$, (3) $A$ prefers $y$ to $x$, and $B$ prefers $x$ to $y$, (4) both $A$ and $B$ prefer $y$ to $x$. 

In a typical market design framework, participants would know the exact mechanism and how it translates each of the four potential problems to a unique outcome. By contrast, in our framework, announcements can be opaque, meaning that for some problems, the description allows a non-singleton set of possible outcomes. Concretely, suppose that in the simple problem described above, the announcement is: ``Our algorithm maximizes the number of individuals who obtain their highest-ranked object." Then, the set of possible outcomes are the following:
\begin{enumerate}
    \item if both $A$ and $B$ prefer $x$ to $y$, then the unique possible outcome is $x$,
    \item if $A$ prefers $x$ to $y$, and $B$ prefers $y$ to $x$, then the set of possible outcomes is $\{x,y\}$,
    \item if $A$ prefers $y$ to $x$, and $B$ prefers $x$ to $y$, then the set of possible outcomes is $\{x,y\}$,
    \item if both $A$ and $B$ prefer $y$ to $x$, then the unique possible outcome is $y$. 
\end{enumerate}
We now entertain the possibility that any mechanism that is consistent with the announcement may be induced and then ask whether strategy-proofness can be guaranteed for each of them. 

Specifically, there are four possible mechanisms corresponding to the different selections of possible outcomes at problems (2) and (3).  Let us call the four mechanisms $(x-x), (x-y), (y-x)$, and $(y-y)$. Here, $(x-x)$ denotes the mechanism that chooses $x$ at both problems (2) and (3), $(x-y)$ denotes the mechanism that chooses $x$ at problem (2) and chooses $y$ at problem (3), and so on. One can now verify that all four possible mechanisms are strategy-proof. In fact, a closer investigation reveals that $(x-x)$ is a \textit{Top Trading Cycles (TTC)} mechanism \citep{shapley/scarf:74} where $A$ owns object $1$ and $B$ owns object $2$, $(x-y)$ is a \textit{Serial Dictatorship (SD)} \citep{satterthwaite1981strategy, svensson:99} where $A$ is the first dictator, $(y-x)$ is a $SD$ where $B$ is the first dictator, and $(y-y)$ is a $TTC$ where $A$ owns $2$ and $B$ owns $1$. Hence, in this example, we have a robust guarantee of strategy-proofness. Yet, the fact that we have only two outcomes plays a crucial role in this possibility.

This is shown in our main impossibility result: In a setting with at least three outcomes, strategy-proofness cannot be guaranteed for \textit{any} level of opacity (Theorem \ref{thm:impossibility-general}). The result highlights the importance of full transparency for robust guarantees of strategy-proofness in real-life social choice problems. We also show that \textit{weak Maskin monotonicity} \citep{kojima/manea:10} can be guaranteed under opaque announcements (Theorems \ref{thm:weak-maskin-one-individual}). The same holds for the well-known invariance property \textit{non-bossiness} \citep{satterthwaite1981strategy}, as established in Theorem \ref{theorem:non-bosiness-general}.

The paper is organized as follows: In section \ref{sec:model}, we introduce the basic opacity framework. In section \ref{sec:mainresult}, we present our main result. Section \ref{sec:otherproperties} provides the possibility results for other properties. Section \ref{sec:relatedliterature} discusses the relation to the literature. Section \ref{sec:conclusion} concludes.

\section{Framework}\label{sec:model}
There is a finite set of individuals $I$ and a finite set of outcomes $X$ where $|X|=N \geq 3$. Each individual $i \in I$ has a weak preference ranking $R_i$, a complete and transitive binary relation over $X$. 
For two outcomes $x,y \in X$ and an individual $i \in I$, we use $x P_i y$ to denote $x R_i y$ and $\neg y R_i x$. 

Let $\mathcal{R}_X$ be the space of all possible preference profiles $R = (R_i)_{i \in I}$ over $X$ that satisfy the following condition: for any $R \in \mathcal{R}_X$ and any $x,y \in X, x \neq y$, there is an $i \in I$ such that $x P_i y$ or $y P_i x$. In other words, we assume that there are no two outcomes $x$ and $y$ such that every individual is indifferent between $x$ and $y$.\footnote{This is without loss of generality since, in the case of a complete indifference between $x$ and $y$, we can relabel $x$ and $y$, and treat them as a single outcome. In that case, opacity would mean that there should be at least two \textit{distinct} possible outcomes at some problem. In this scenario our results would carry through.} 

We say that a subset $\mathcal{R} \subseteq \mathcal{R}_X$ is \textit{rich}, if whenever there is a preference profile $R\in \mathcal{R}$, an individual $i \in I$, and two outcomes $x,y \in X$, such that $x P_i y$, then there is another preference ranking $R_i' \neq R_i$ of $i$, such that $(R_i',R_{-i}) \in \mathcal{R}$, and $x P_i z \implies x P_i' z, \forall z \in X$. 

An \textit{environment} is a triplet $(I,X,\mathcal{R})$, where $\mathcal{R} \subseteq \mathcal{R}_X$ is an arbitrary rich subset of preference profiles.\footnote{We require a richness assumption to exclude the possibility of trivial settings. For example, when $\mathcal{R}$ is a singleton, strategy-proofness and all other studied properties hold trivially. Also, note that all studied social choice environments with at least three alternatives satisfy our richness condition.} We refer to an element $R \in \mathcal{R}$ as a problem. Note that this model covers a wide range of applications, including the classic social choice setup \citep{arrow:51}, matching with contracts \citep{hatfield/milgrom:05}, and constrained object allocation \citep{root/ahn:23}, among many others.\footnote{For example, in the matching with contracts environment \citep{hatfield/milgrom:05}, $\mathcal{R}$ corresponds to all preferences profiles where every agent is indifferent between two outcomes that give them the same match and the same contract. Matching with contracts covers various applications, such as object allocation without money or auctions with discrete-valued bids.}

The following definitions will be the building blocks of our framework.
Fix an arbitrary environment $(I,X,\mathcal{R})$. An \textbf{announcement} $\Pi: \mathcal{R} \mapsto 2^{X} \setminus \{\emptyset\}$ is a correspondence that describes the set of possible outcomes that can be induced. A (direct) mechanism is a mapping $\varphi: \mathcal{R} \rightarrow X$ that specifies an outcome for each problem. Given an announcement $\Pi$, a \textbf{possible mechanism} is a mechanism $\varphi$ such that $\varphi(R) \in \Pi(R)$ for all $R \in \mathcal{R}$. Let $\Phi(\Pi)$ denote the set of all possible mechanisms given $\Pi$. We are ready to define what we mean by opacity in this setup.

We say that an announcement $\Pi: \mathcal{R} \mapsto 2^{X} \setminus \{\emptyset\}$ is \textbf{fully transparent} if $\Pi(R)$ is a singleton for every $R \in \mathcal{R}$. If $\Pi$ is not fully transparent, then we say that $\Pi$ is \textbf{opaque}.  If $\Pi$ is fully transparent, then $\Phi(\Pi)$ contains a single mechanism. Conversely, if an announcement $\Pi$ is opaque, then $|\Phi(\Pi)| > 1$. We say that an announcement $\Pi$ \textbf{guarantees} a property if every $\varphi \in \Phi(\Pi)$ satisfies the property.  
The following property will be central to our analysis. A mechanism $\varphi: \mathcal{R} \rightarrow X$ is \textbf{strategy-proof} if for every $i \in I$, and every $R, R' \in \mathcal{R}$, where $R'$ differs from $R$ only with respect to the preference ranking of $i$, we have that $\varphi(R) R_i \varphi(R')$.

In the rest of this work, we study whether some desirable properties of mechanisms can be guaranteed under opaque announcements.

\section{Main Result}\label{sec:mainresult}

This section presents our main result, which shows that strategy-proofness cannot be guaranteed under any opaque announcement.  

\begin{theorem}
\label{thm:impossibility-general}
Let $(I,X,\mathcal{R})$ be an arbitrary environment. 
An announcement $\Pi$ guarantees strategy-proofness if and only if $\Pi$ is fully transparent, and the unique possible mechanism given $\Pi$ is strategy-proof.
\end{theorem}

\begin{proof}

The `if' direction is trivial. In what follows, we prove the `only if' direction. 

Suppose that an announcement $\Pi$ guarantees strategy-proofness. 
If $\Pi$ is fully transparent, and the unique possible mechanism given $\Pi$ is not strategy-proof, then $\Pi$ does not guarantee strategy-proofness. Therefore, if $\Pi$ is fully transparent, the unique possible mechanism should be strategy-proof. Hence, to prove the `only if' direction, it is left to show that when $\Pi$ guarantees strategy-proofness, it is fully transparent.  

The proof is by contraposition. We will assume that $\Pi$ is opaque and show that $\Pi$ does not guarantee strategy-proofness. 

Since $\Pi$ is opaque, there is a problem $R$, such that $\Pi(R)$ is non-singleton. Let $x$ and $y$ be two different outcomes in $\Pi(R)$. Let $i$ be an individual who is not indifferent between $x$ and $y$. Such an individual exists because we assumed that there are no two outcomes towards which all individuals are indifferent. 
Without loss of generality, suppose that $x P_i y$. Consider two possible mechanisms $\varphi, \varphi' \in \Phi(\Pi)$ that only differ from each other by that $\varphi(R) = x$ and $\varphi'(R) = y$, and they agree for all other problems, i.e., for all $R' \neq R$, $\varphi(R') = \varphi'(R') $. 
To prove that $\Pi$ does not guarantee strategy-proofness, it is sufficient to show that at least one of the mechanisms in $\{\varphi, \varphi'\}$ is not strategy-proof. 
Without loss of generality, we will assume that $\varphi'$ is  strategy-proof, and we will show that $\varphi$ cannot be strategy-proof. 

Consider an arbitrary preference ranking $R' \neq R$ that differs from $R$ only with respect to the ranking of individual $i$, and it does not rank any new objects weakly above $x$. That is, for all $z \in X$, $x P_i z \implies  x P_i' z$. By our richness assumption, such an $R'$ exists.

Since $\varphi'$ is strategy-proof, and $x P_i y = \varphi'(R)$, it should be that $x P_i \varphi'(R')$. 

Because $x P_i \varphi'(R')$, and by our choice of $R'$, we have that $x P_i' \varphi'(R')$. Moreover, since $\varphi$ and $\varphi'$ agree on all problems other than $R$, we have that $\varphi'(R') = \varphi(R')$. Thus, \[ \varphi(R) = x P_i' \varphi'(R') = \varphi(R'),\]
which implies that $\varphi$ is not strategy-proof. This completes the proof of Theorem \ref{thm:impossibility-general}. \end{proof}

\section{Other Properties
}\label{sec:otherproperties}

This section shows that some monotonicity and invariance properties can be guaranteed under opaque announcements. 

We start with \textbf{weak Maskin monotonicity} \citep{kojima/manea:10}. We say a problem $R' \in \mathcal{R}$ is a \textit{monotonic transformation} of problem $R \in \mathcal{R}$ at outcome $x \in X$, if for all $i \in I$ and all $y \in X$, we have that $x P_i y \implies x P_i' y$. 
We say a mechanism $\varphi: \mathcal{R} \rightarrow X$ is \textbf{weakly Maskin monotonic}, if for any two problems $R,R' \in \mathcal{R}$, if $R'$ is a monotonic transformation of $R$ at outcome $\varphi(R)$, then $\varphi(R') R_i' \varphi(R)$ for all $i \in I$. 

\begin{theorem}
\label{thm:weak-maskin-one-individual}
For some environments, there are opaque announcements that guarantee weak Maskin monotonicity.
\end{theorem}

\begin{proof}
Consider an environment with a single individual and strict preferences. Since the preferences are strict for this individual, we will refer to a generic problem with $P$ rather than $R$. 

Suppose the outcomes are indexed, i.e., $X = \{x_1, x_2, \dots, x_N\}$. Throughout the proof, let $\bar{P}$ denote the preference ranking with $x_1 \bar{P} x_2 \bar{P} \dots \bar{P} x_N$.

Take any $P$, such that $P \neq \bar{P}$. Then, at least one object changes its rank under $P$ relative to $\bar{P}$. We introduce the following notation. Let $X(P)$ be the set of outcomes that strictly improve their rank at $P$ compared to $\bar{P}$, that is, 
\[
X(P) = \Big\{ x \in X : \big| \big\{ y \in X : y P x \big\} \big| < \big| \big\{ y \in X : y \bar{P} x \big\} \big| \Big\}.
\]
Let $x(P)$ be the highest ranked outcome according to $P$ at $X(P)$, that is, 
\[
x(P) = \argmin_{x \in X(P)} \big| \big\{ y \in X : x P y \big\} \big|.\footnote{For example, if $x_1 P x_5 P x_4 P x_2 P x_3$, then $X(P) = \{x_5, x_4\}$ and $x(P) = x_5$.}
\]

Our goal is to construct an opaque announcement that guarantees weak Maskin monotonicity.
 
Consider the opaque announcement $\Pi: \mathcal{R} \mapsto 2^{X} \setminus \emptyset$, where
\begin{itemize}
    \item for $P = \bar{P}$, we set $\Pi(P) = \{x_{N-1}, x_N\}$ and,
    \item for any $P \neq \bar{P}$, we set $\Pi(P) = \{ x(P) \}$.
\end{itemize}

This implies there are only two possible mechanisms given $\Pi$, i.e., $\Phi(\Pi) = \{\varphi,\psi\}$. Moreover, these two mechanisms differ only with respect to the outcome they choose at $\bar{P}$. Specifically, let $\varphi$ denote the mechanism with $\varphi(\bar{P}) = x_{N-1}$ and let $\psi$ denote the mechanism with $\psi(\bar{P}) = x_N$. To finish the prove, we must establish that both $\varphi$ and $\psi$ are weakly Maskin monotonic.

Here, we only prove the weak Maskin monotonicity of $\varphi$. The proof for weak Maskin monotonicity of $\psi$ is almost identical and therefore omitted.

Take any pair $P,P' \in \mathcal{R}$. 
To show that $\varphi$ is weakly Maskin monotonic, we consider two cases.

\textbf{Case 1.}  $P \neq \bar{P}$ and $P' \neq \bar{P}$.

Without loss of generality, suppose that $P'$ is a monotonic of transformation of $P$ at outcome $\varphi(P)$. 

Let $x_n = \varphi(P) = x(P)$. By definition of $x(P)$, it should be that either (i) $x_n \neq x_1$ and $x_n$ is the highest ranked outcome according to $P$, or (ii) there is an index $m < n-1$, such that for all $m' < m'' \leq m$, we have that $x_{m'} P x_{m''} P x_n$, and for all $m''' > m$, we have that $x_n P x_{m'''}$. To put it simply, in the latter case, the preference ranking $P$ is of the form $x_1 P x_2 \dots P x_{m-1} P x_m P x_n \dots$ for some $m < n-1$.

We consider cases (i) and (ii) separately.

(i) Suppose $x_n \neq x_1$ and $x_n$ is the highest ranked outcome according to $P$. Since $P'$ is a monotonic transformation of $P$ at outcome $x_n$, it should be that $x_n$ is also the highest ranked outcome according to $P'$. Hence, by definition, $x(P') = x_n$, which is trivially weakly more preferred than $x_n$ according to $P'$.

(ii) Suppose there is an index $m < n-1$, such that for all $m' < m'' \leq m$, we have that $x_{m'} P x_{m''} P x_n$, and for all $m''' > m$, we have that $x_n P x_{m'''}$.

Since $P'$ is a monotonic transformation of $P$, it should be that no outcome with index strictly larger than $m$ is ranked higher than $x_n$ under $P'$. Consider the set $X(P')$. By the previous observation, $x_n$ can only improve its rank when we go from $P$ to $P'$, and therefore $x_n \in X(P)$ implies that $x_n \in X(P')$. By definition, $\varphi(P') = x(P')$ is the highest ranked outcome at $X(P')$ according to $P'$, and therefore, it is weakly more preferred than $x_n$ according to $P'$.

\textbf{Case 2.} Consider two problems $P$ and $P'$, such that $P =  \bar{P}$ and $P' \neq \bar{P}$. 

We first show that $P = \bar{P}$ cannot be a monotonic transformation of $P'$ at $\varphi(P')$. 

Let $x_n = \varphi(P') = x(P')$. By definition of $x(P')$, we have that $x_n$ is ranked at a strictly higher position at $P'$ than under $\bar{P}$. Therefore, it is immediate that $\bar{P}$ is not a monotonic transformation of $P'$ at outcome $x_n$. 

Finally, suppose that $P'$ is a monotonic transformation of $P = \bar{P}$ at $\varphi(\bar{P})$. By definition of $\varphi$, we have that $\varphi(\bar{P}) = x_{N-1}$. 

Let $x_n = \varphi(P') = x(P')$. 
By definition of $x(P')$, it should be that either (i) $x_n \neq x_1$ and $x_n$ is the highest ranked outcome according to $P'$, or (ii) there is an index $m < n-1$, such that for all $m' < m'' \leq m$, we have that $x_{m'} P' x_{m''} P' x_n$, and for all $m''' > m$, we have that $x_n P' x_{m'''}$, that is, the preference ranking $P'$ is of the form $x_1 P' x_2 \dots P' x_{m-1} P' x_m P' x_n \dots$ for some $m < n-1$.

In either case, it is immediate that $x_n$ is weakly more preferred than $x_{N-1}$ according to $P'$. (Note that it can also be that $x_n = x_{N-1}$). This completes the proof of Theorem \ref{thm:weak-maskin-one-individual}. \end{proof}

Finally, we adapt non-bossiness as defined by \cite{satterthwaite1981strategy} to our general setup. A mechanism $\varphi$ is \textbf{non-bossy}, if for any two problems $R, R' \in \mathcal{R}$, that only differ with respect to the preferences of an 
individual $i \in I$, it should be that whenever $\varphi(R) P_j \varphi(R')$ or $\varphi(R') P_j \varphi(R)$ for some $j \neq i$, then $\varphi(R) P_i \varphi(R')$ or $\varphi(R') P_i \varphi(R)$. In other words, non-bossiness ensures that by changing her preference report, an individual cannot harm or benefit others without harming or benefiting herself.\footnote{In the object allocation setup with strict preferences, our definition boils down to the non-bossiness condition by \cite{satterthwaite1981strategy} and \cite{svensson:99}.}

\begin{theorem}
\label{theorem:non-bosiness-general}
For some environments, there are opaque announcements that guarantee non-bossiness.
\end{theorem}

\begin{proof}
First, note that non-bossiness holds for any environment with a single individual. In what follows, we prove a stronger statement. Namely, we show that opaque announcements can guarantee non-bossiness in environments with any number of individuals. 
Consider an environment $(I,X,\mathcal{R})$ that satisfies the following condition: there are some two outcomes $x,y \in X$, such that for any $R \in \mathcal{R}$ and for any $i \in I$, we have that $x P_i y$ or $y P_i x$.\footnote{For example, the object allocation setup with strict preferences satisfies this condition. There, $x$ and $y$ need to be two allocations that assign different objects to all individuals.}

Consider an arbitrary opaque announcement $\Pi: \mathcal{R} \mapsto 2^{X} \setminus \emptyset$, such that $\Pi(P) \subseteq \{x,y\}$ for all $P \in \mathcal{R}$. Then, by construction, $\Pi$ is non-bossy.  \end{proof}

\section{Related Literature}\label{sec:relatedliterature}

To the best of our knowledge, ours is the first paper to formalize opacity and its conflict with guaranteeing other normative properties in social choice and market design. In that regard, our research question and results have no close counterpart in the existing economics literature.

Our work is related to the mechanism design literature on partial or limited commitment. Given a partial description of a mechanism, the designer or central planner cannot commit to a particular selection rule and can vary the selection of the mechanism based on agents' type reports. Unlike the mechanism design literature on limited commitment \citep{bester2000imperfect,bester2001contracting,baliga1997theory,doval2022mechanism}, we do not study central planner's preferences and credible selections. Instead, we take a robust approach where the agents entertain the possibility of every mechanism selection by the planner. Our main question is different from those in all these other works; we are interested in guaranteeing a property of a mechanism in a robust sense, namely, that every possible selection from a set of mechanisms satisfies the property.

Other notions of opacity have been studied in some economic environments. \cite{haupt2023opaque} consider opacity of contractual terms in a moral hazard model. \cite{stahl2017certification} examine some form of opacity in the context of product quality with and without third-party certification. In a recent empirical paper, \cite{Kapor2024} investigates the role of opacity in college admission criteria for different applicant types.

More broadly, our paper contributes to the literature on transparency in allocation and social choice problems \citep{akbarpour2020credible, grigoryan/moller:2024aea, grigoryan/moller:23, moller2022transparent,hakimov2023improving, li2017obviously}. There is a publicly announced and fully known mechanism in these papers, and there are concerns that the designer or central planner may  \textit{deviate} from the mechanism subject to not being detected by the participants. In our setup, the announcement is not a mechanism; hence, our paper captures a different dimension of non-transparency in these problems. 

Finally, our work is also related to the literature on simplicity in market design and matching \citep{li2017obviously, pycia2023theory}. These papers analyze incentives when agents have a limited understanding of the mechanism's working. 
Interestingly, the variants of strategy-proofness considered in these papers are compatible with these frameworks' respective forms of uncertainty and partial commitment. By contrast, our robust notion of strategy-proofness fails for any opacity level.

\section{Conclusion}\label{sec:conclusion}

We introduce a framework of opacity in social choice and market design and show that strategy-proofness cannot be robustly guaranteed whenever there is any level of opacity. Hence, in our framework, full transparency is necessary to ensure the optimality of participants' truth-telling. We also show that some well-known monotonicity and invariance properties are compatible with opacity. We have established this compatibility in specific environments but not generally. As a potential future research question, it will be interesting to understand the strongest incentive-, monotonicity-, or invariance properties that can be guaranteed under opacity in general environments. For now, we leave this question for future research.

\bibliography{bibmatching}

\end{document}